\documentclass[superscriptaddress,showpacs,twocolumn,prb]{revtex4}
\usepackage{amsfonts}
\usepackage{amsmath}
\usepackage{amssymb}
\usepackage{graphicx}

\begin{document}

\title{Spin edge states: an exact solution and oscillations of the spin
current}
\author{V. L. Grigoryan}
\affiliation{Department of Radiophysics, Yerevan State University, 1 A. Manoukian St.,
375025 Yerevan, Armenia}
\author{A. Matos Abiague}
\affiliation{Department of Physics, University of Regensburg, 93040 Regensburg, Germany}
\author{S. M. Badalyan}
\email{Samvel.Badalyan@physik.uni-regensburg.de}
\affiliation{Department of Radiophysics, Yerevan State University, 1 A. Manoukian St.,
375025 Yerevan, Armenia}
\affiliation{Department of Physics, University of Regensburg, 93040 Regensburg, Germany}
\date{\today }

\begin{abstract}
We study the spin edge states, induced by the combined effect of spin-orbit
interaction (SOI) and hard-wall confining potential, in a two-dimensional
electron system, exposed to a perpendicular magnetic field. We find an exact
solution of the problem and show that the spin resolved edge states are
separated in space. The SOI generated rearrangement of the spectrum results
in a peaked behavior of the net spin current versus the Fermi energy. The
predicted oscillations of the spin current with a period, determined by the
SOI renormalized cyclotron energy, can serve as an effective tool for
controlling the spin motion in spintronic devices.
\end{abstract}

\pacs{72.25.Dc, 73.21.Fg, 73.63.Hs, 72.10.-d}
\maketitle

\section{Introduction}

The principal importance of SOI is in its ability to link the electron
charge and spin degrees of freedom, which is fertile for novel physical
phenomena~\cite{fmesz,zfs,wolf}. Unlike the charge, the electron spin is
double-valued and identifies two system components, which can be separated
as in the spin Hall effect~\cite{dyakonov,kato} or mixed via the spin
Coulomb drag~\cite{weber1,smb1}. There are different mechanisms, realizing
SOI~\cite{fmesz}, and the interplay between them produces another rich arena
for study and potential applications in spintronics~\cite{bernevig,smb2}.


In two dimensional electron systems (2DES) of the quantum Hall effect
geometry, the extended edge state play a central role in understanding of
transport phenomena~\cite{laughlin,halperin,streda}. The suppression of
backscattering 
and inter-edge-state relaxation~\cite{smb6,muller,khaetskii} make possible
non-dissipative transport through edge channels. In the presence of SOI the
nonlocal transport through spin-polarized edge channels holds promise of
providing even richer phenomenology and greater power of electronic
applications. Another strong motivation for investigations of spin edge
states is related to the recent experimental realization of the Mach-Zehnder~%
\cite{mahalu}.

Recently several theoretical papers have addressed the effect of SOI on the
edge states in restricted 2DES~\cite{pala,bao,kramer,balseiro,wang} and
along magnetic interfaces~\cite{ngo,su}. All these works, however, find
unlikely an exact analytical solution of the edge state problem and adopt
different numerical approaches~\cite{bao,balseiro,wang,ngo,su}, use a
parabolic confining potential~\cite{kramer}, which has no flat domain, or
give an analytical approximation in the limit of strong magnetic fields~\cite%
{pala} where the effective SOI coupling is small.

Here we present an analytical solution to the spin edge states, induced by
the combined effect of SOI and hard-wall confining potential. 
We derive an \textit{exact} formula for the electron energy dispersions and
calculate their spectral and transport properties. We find that due to SOI
the spin edge states are resolved not only in the energy but are also 
\textit{separated in space}: an effect, which is not captured by the
approximate approach, adopted in Ref.~\onlinecite{pala}. From the energy
spectrum we calculate the electron group velocity and the average spin
components. We find that the magnitude of spin components are not equal in
the up and down states. In the quasibulk states the electron spins are
mainly aligned along the magnetic field while near the hard-wall the spins
of edge states become mainly perpendicular to their propagation direction.
Using these ingredients we calculate the components of the net spin current
as a function of the Fermi energy and show that the SOI induced splitting
results in the peaked behavior of the spin current. We argue that the
predicted oscillations of the spin current with a period, determined by the
renormalized cyclotron frequency, can serve as a new tool for manipulating
spin currents in a controllable manner. The developed approach here is
equally applicable to the spin edge states along magnetic interfaces in
nonhomogeneous magnetic fields.

\section{Theoretical concept}

We assume that the 2DES resides in a quantum well, formed in the (001) plane
of a zincblende semiconductor heterostructure, and is exposed to a
perpendicular homogeneous magnetic field, $\vec{B}=B_{0}\hat{\mathbf{z}}$.
The motion of electrons in the 2DES is confined by an infinite hard-wall
potential, $V(x)=\infty $ for $x<0$. Such a system is described by a
two-dimensional Hamiltonian of the form 
\begin{equation}
H=H_{0}+H_{SOI}+V(x)  \label{eqn1}
\end{equation}%
where the Hamiltonian of free particle in a magnetic field is $H_{0}=\left( 
\vec{\pi}^{2}/2m^{\ast }\right) \hat{\tau}$ and the Rashba SOI Hamiltonian $%
H_{SOI}=\alpha _{R}\left( \pi _{x}\hat{\sigma}_{y}-\pi _{y} \hat{\sigma}%
_{x}\right) $~\cite{rashba}, $m^{\ast }$ denotes the electron effective
mass, $\vec{\pi}=\vec{p}-(e/c)\vec{A}$ the kinetic momentum with $\vec{p}%
=-i\hbar \vec{\nabla}$. We choose the Landau gauge so that the components of
vector potential are $\vec{A}=\left( 0,xB_{0},0\right) $. The unity matrix $%
\hat{\tau}$ and the Pauli spin matrices $\hat{\sigma}$ act in the spin
space. It is assumed that electrons are confined to the lowest energy
subband in the $z$-direction.

Using the \textit{ansatz }$\Psi \left( x,y\right) =e^{ik_{y}y}\chi
_{k_{y}}(x)$, we can reduce the two-dimensional Schr\"{o}dinger equation $%
H\Psi =E\Psi $ to the one-dimensional problem where $E$ is the electron
total energy and $k_{y}$ the electron momentum in $y-$direction.%
Then the electron wave function $\chi _{k_{y}}(x)$ in $x-$direction should
satisfy the following equation 
\begin{gather}
\left\{ \left[ \frac{d^{2}}{dx^{2}}+\nu +\frac{1}{2}-\frac{\left( x-X\left(
k_{y}\right) \right) ^{2}}{4}\right] \hat{\tau}\right.  \label{eqn6} \\
\left. +\gamma \left[ i\frac{d}{dx}\hat{\sigma}_{y}-\frac{x-X(k_{y})}{2}\hat{%
\sigma}_{x}\right] \right\} \chi _{k_{y}}\left( x\right) =0~,  \notag
\end{gather}%
where the effective potential $V_{\text{eff}}(x)=\left( x\text{ }-X\left(
k_{y}\right) \right) ^{2}/4$ in $x-$direction depends on the wave vector $%
k_{y}$. In Eq.~\ref{eqn6} we express the energy $E\rightarrow \left( \nu
+1/2\right) \hbar \omega _{B}$ in units of the cyclotron energy, $\hbar
\omega _{B}\equiv \hbar eB_{0}/m^{\ast }c$, and the length $x\rightarrow
xl_{B}/\sqrt{2}$ in the magnetic length, $l_{B}\equiv \sqrt{\hbar c/eB_{0}}$%
. We introduce also the dimensionless SOI coupling constant $\gamma =\sqrt{2}%
\alpha _{R}/v_{B}$ with the cyclotron velocity $v_{B}=\hbar ^{2}/m^{\ast
}l_{B}$ and the dimensionless coordinate of the center of orbital rotation $%
X(k_{y})=\sqrt{2}k_{y}l_{B}$. 
\begin{figure*}[t]
\centering
\includegraphics[width=.32\linewidth]{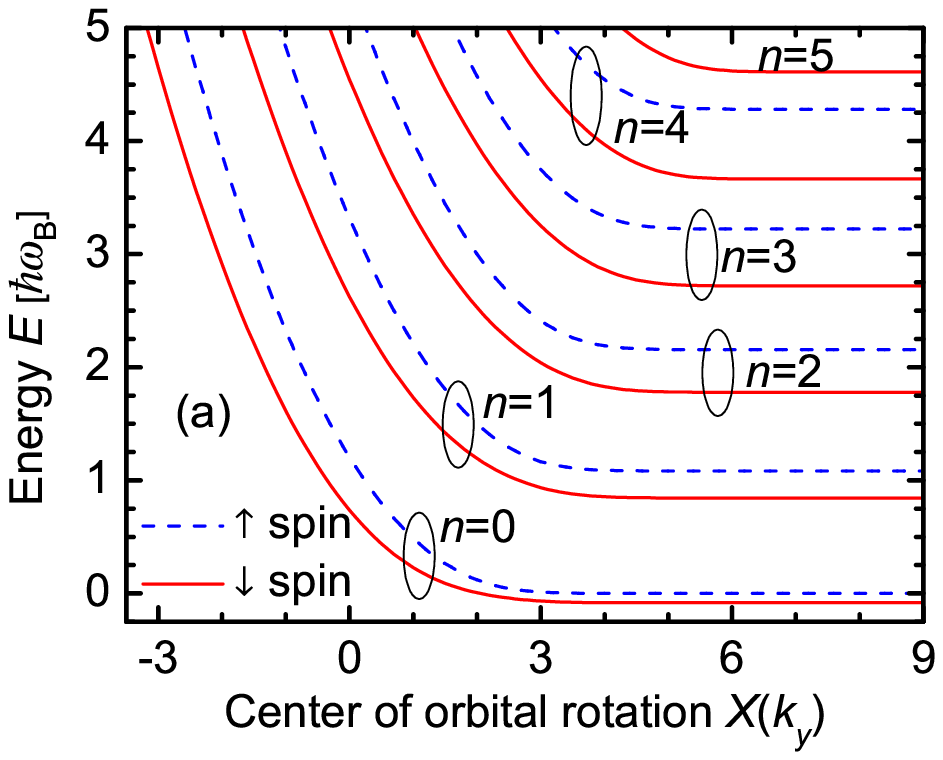} \hfill %
\includegraphics[width=.32\linewidth]{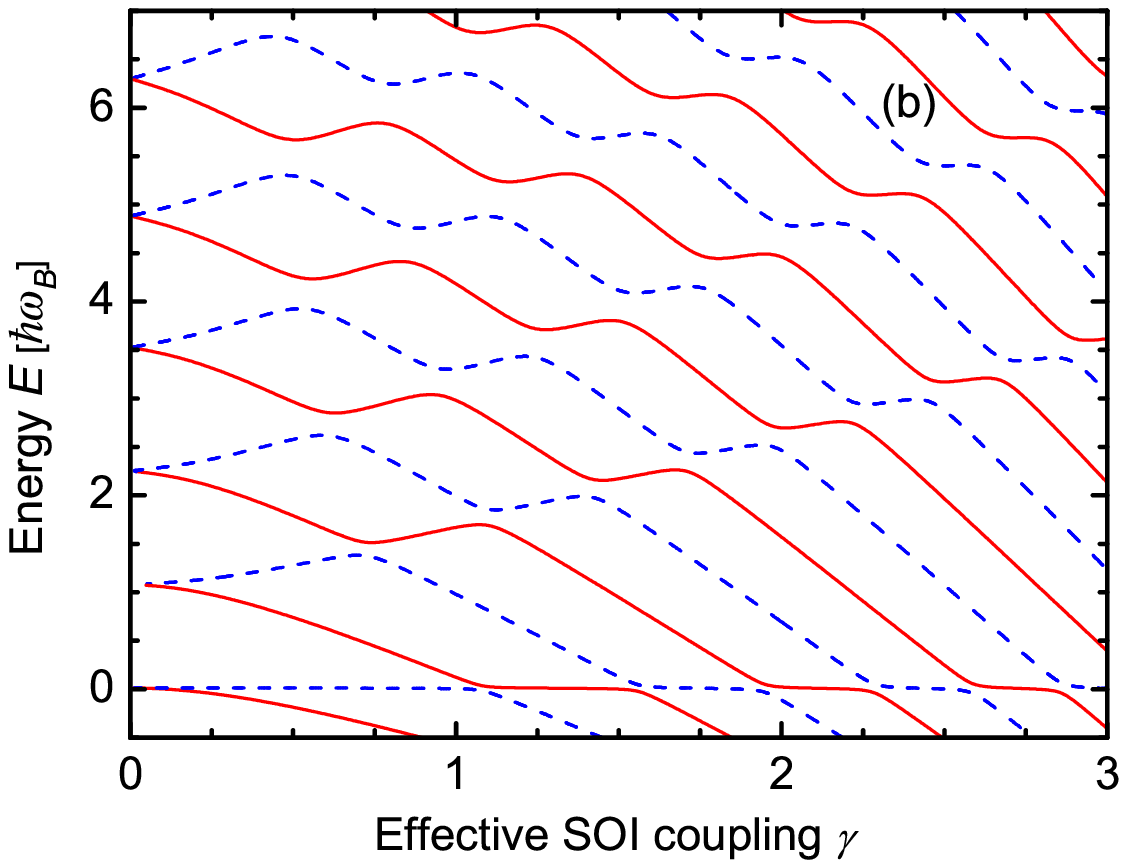} \hfill %
\includegraphics[width=.32\linewidth]{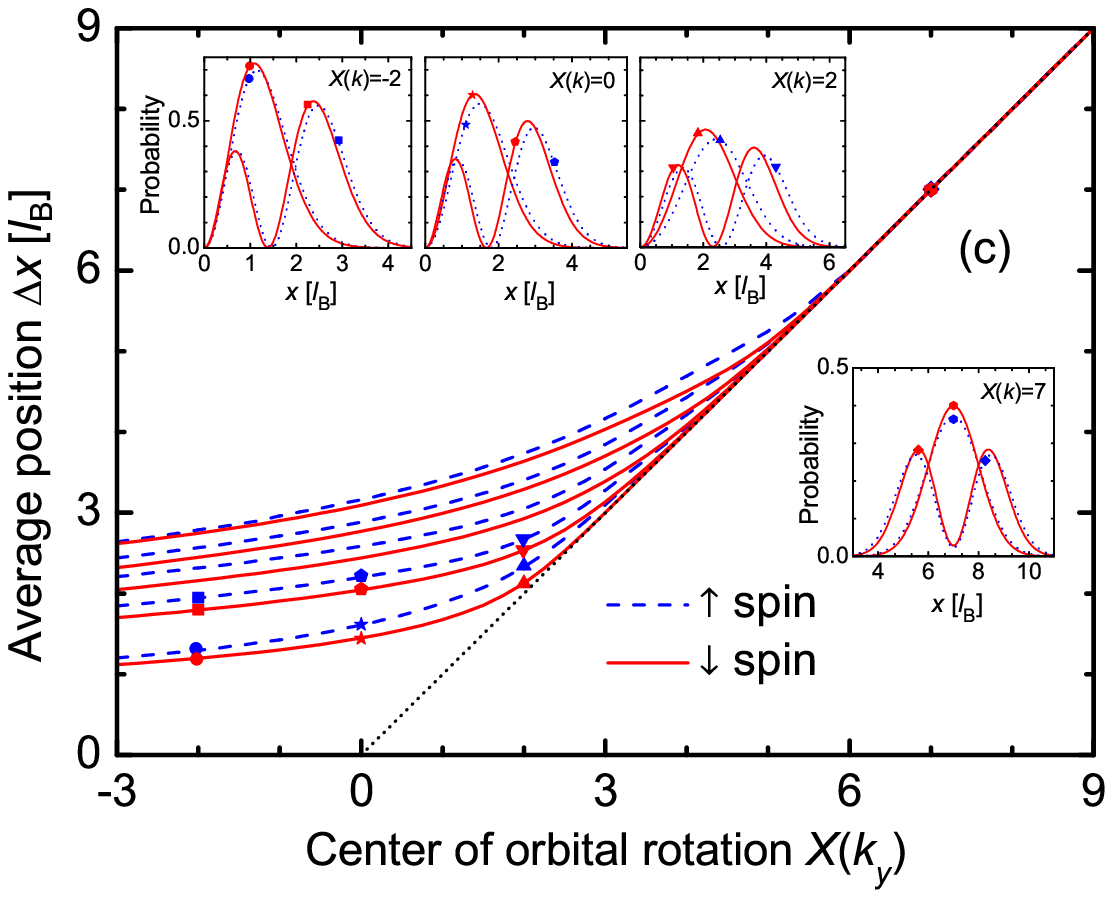}
\caption{(Color online) The energy spectrum of spin edge states as a
function (a) of the momentum $k_{y}$ for $\protect\gamma =0.3$ and (b) of
the effective SOI coupling $\protect\gamma $ (or that is the same as a
function of $B^{-1/2}$) for $X(k_{y})=3$. The dashed and solid curves
correspond to the up and down spins. (c) The particle average position as a
function of $k_{y}$ for $\protect\gamma =0.3$. The probability density in
the respective symbol positions is shown in the insets.}
\label{fig1}
\end{figure*}
Taking into account explicitly that the eigenstates of Eq.~\ref{eqn6} are
spinors, 
\begin{equation}
\chi _{k_{y}}\left( x\right) =\left\vert 
\begin{array}{c}
\chi _{1k_{y}}\left( x-X(k_{y})\right) \\ 
\chi _{2k_{y}}\left( x-X(k_{y})\right)%
\end{array}%
\right\vert ~,  \label{eqn7}
\end{equation}%
we write the Schr\"{o}dinger equation in the following compact form 
\begin{equation}
\left( 
\begin{array}{cc}
h_{\nu } & h_{+} \\ 
h_{-} & h_{\nu }%
\end{array}%
\right) 
\begin{pmatrix}
\chi _{1k_{y}}\left( \xi \right) \\ 
\chi _{2k_{y}}\left( \xi \right)%
\end{pmatrix}%
=0~.  \label{eqn8}
\end{equation}%
Here we introduce the following operators: 
\begin{align}
h_{\nu }& =\left( \frac{d^{2}}{d\xi ^{2}}+\nu +\frac{1}{2}-\frac{\xi ^{2}}{4}%
\right) ~,  \label{eqn9} \\
h_{\pm }& =-\gamma \left[ \frac{\xi }{2}\mp \frac{d}{d\xi }\right] ~.
\label{eqn10}
\end{align}%
The system of equations, obtained from Eq.~\ref{eqn8} has to be solved under
the boundary conditions $\chi _{k_{y}}\left( x\right) \rightarrow 0$ when $%
x\rightarrow 0$ and $+\infty $. In the absence of SOI, $h_{\pm }=0$, the
solution is given in terms of the parabolic cylindrical functions, $D_{\nu
}(x)$. In the presence of SOI we search the bulk solution of matrix equation
(\ref{eqn8}) as $\chi _{1k_{y}}\left( \xi \right) =aD_{\mu }\left( \xi
\right) $ and $\chi _{2k_{y}}\left( \xi \right) =bD_{\mu -1}\left( \xi
\right) $ where $a$ and $b$ are the $x$ independent spinor coefficients and $%
\mu $ is an arbitrary index, different from $\nu $. Making use of the
following recurrent properties of the parabolic cylindrical functions 
\begin{align}
h_{\nu }D_{\mu }\left( \xi \right) & =\left( \nu -\mu \right) D_{\mu }\left(
\xi \right) ~,  \label{eqn13} \\
h_{\pm }D_{\mu }\left( \xi \right) & =-\gamma \left\{ 
\begin{array}{c}
D_{\mu +1}\left( \xi \right) \\ 
\mu D_{\mu -1}\left( \xi \right)%
\end{array}%
\right. ~,  \label{eqn133}
\end{align}%
we obtain%
\begin{align}
\mu _{\pm }\left( \nu ,\gamma \right) & =\nu +\frac{1}{2}+\frac{\gamma ^{2}}{%
2}\pm \sqrt{\nu \gamma ^{2}+\frac{1}{4}\left( 1+\gamma ^{2}\right) ^{2}}~,
\label{eqn15} \\
c_{\pm }\left( \nu ,\gamma \right) & =-\frac{1}{\gamma }\left( \frac{1}{2}+%
\frac{\gamma ^{2}}{2}\pm \sqrt{\nu \gamma ^{2}+\frac{1}{4}\left( 1+\gamma
^{2}\right) ^{2}}\right)  \label{eqn16}
\end{align}%
where $c_{\pm }=b_{\pm }/a_{\pm }$. Thus, the two independent bulk solutions
of Eq.~\ref{eqn8} are given by the spinor wave functions%
\begin{equation}
\chi _{k_{y}}^{\pm }(\xi )=a_{\pm }\left\vert 
\begin{array}{c}
D_{\mu _{\pm }\left( \nu ,\gamma \right) }\left( \xi \right) \\ 
c_{\pm }D_{\mu \pm \left( \nu ,\gamma \right) -1}\left( \xi \right)%
\end{array}%
\right\vert ~.  \label{eqn17}
\end{equation}%
The normalization of the wave functions $\int d\xi \chi _{k_{y}}^{\dag }(\xi
)\chi _{k_{y}}(\xi )=1$ gives the amplitude of eigenstates

\begin{equation}
a_{\pm }=\left[ \int d\xi \left( \left\vert D_{\mu _{\pm }}(\xi )\right\vert
^{2}+\left\vert c_{\pm }\right\vert ^{2}\left\vert D_{\mu _{\pm }-1}(\xi
)\right\vert ^{2}\right) \right] ^{-1/2}~.
\end{equation}%
It is easy to see that in the limit of vanishing SOI, $\gamma \rightarrow 0$%
, we have $a_{+}\rightarrow 0$ and $b_{-}\rightarrow 0$ and recover the
usual edge states, which are doubly degenerated with respect to the spin:

\begin{equation}
\chi _{k_{y}}^{+}(x)\sim D_{\nu }\left( x\right) \left\vert 
\begin{array}{c}
1 \\ 
0%
\end{array}%
\right\vert \text{ and }\chi _{k_{y}}^{-}(x)\sim D_{\nu }\left( x\right)
\left\vert 
\begin{array}{c}
0 \\ 
1%
\end{array}%
\right\vert ~.
\end{equation}

On the other hand the solution (\ref{eqn17}) for sufficiently large values
of $X\left( k_{y}\right) $\ describes quasibulk Landau states so that the
index $\mu _{\pm }\left( \nu ,\gamma \right) $\ differs only exponentially
from the Landau index $l=0,1,2\ldots $\ and the parabolic cylindric
functions are given by their asymptotics Hermite polynomials, $D_{l}\left(
\xi \right) =2^{-l/2}\exp \left( -\xi ^{2}/4\right) H_{l}(\sqrt{\xi }/2).$\
In the limit of $X\left( k_{y}\right) \rightarrow \infty $\ taking $\mu
_{\pm }\left( \nu ,\gamma \right) =l$\ in Eqs.$~\left( \ref{eqn15}\right) $\
and $\left( \ref{eqn16}\right) $, one can exactly reproduce the spectrum and
the wave functions of the bulk dispersionless Landau levels, renormalized by
the SOI for $l=1,2\ldots $~\cite{rashba,john,shen1,shen2,lipparini}%
\begin{align}
E_{l}^{\pm }(\gamma )& =\left( l\pm \sqrt{\frac{1}{4}+l\gamma ^{2}}\right)
\hbar \omega _{B}~, \\
c_{\pm }\left( \nu ,\gamma \right) & =-\frac{1}{\gamma }\left( \frac{1}{2}%
\mp \sqrt{\frac{1}{4}+l\gamma ^{2}}\right) .
\end{align}%
As usual the $l=0$\ Landau level remains not perturbed by the spin-orbit
coupling.

In order to obtain the spectrum of the spin edge states we require vanishing
of the electron wave function (\ref{eqn17}) at $x=0$. The energy, $E$,
obtained from the vanishing conditions for both spinor components at $x=0$,
should be the same. As seen, however, the different spinor components of the
bulk solution $(\ref{eqn17})$ are given by the parabolic cylindric functions
with different indices, which makes impossible their vanishing
simultaneously. To satisfy the boundary conditions, we construct a linear
combination of the two independent bulk solutions as 
\begin{equation}
\psi _{k_{y}}(\xi )=\alpha \chi _{k_{y}}^{+}(\xi )+\beta \chi
_{k_{y}}^{-}(\xi )~.  \label{eqn22}
\end{equation}%
and choose the coefficients $\alpha $ and $\beta $\ so that the components
of the new spinor wave function $\psi _{k_{y}}(\xi )$ vanish at $x=0$. The
eigenvalue problem for $\alpha $ and $\beta $\ has a solution if the
respective determinant vanishes at $x=0$. This leads to the following 
\textit{exact} dispersion equation 
\begin{eqnarray}
&&c_{-}D_{\mu _{+}}\left( -X(k_{y})\right) D_{\mu _{-}-1}\left(
-X(k_{y})\right)  \notag \\
&=&c_{+}D_{\mu _{-}}\left( -X(k_{y})\right) D_{\mu _{+}-1}\left(
-X(k_{y})\right)  \label{eqn23}
\end{eqnarray}%
for the spin edge states with the wave functions 
\begin{equation}
\psi _{k_{y}}(\xi )=\alpha \left\vert 
\begin{array}{c}
D_{\mu _{+}}\left( \xi \right) -rD_{\mu _{-}}\left( \xi \right) \\ 
c_{+}D_{\mu _{+}-1}\left( \xi \right) -rc_{-}D_{\mu _{-}-1}\left( \xi \right)%
\end{array}%
\right\vert ~.  \label{eqn24}
\end{equation}%
Here $r=D_{\mu _{+}}\left( -X(k_{y})\right) /D_{\mu _{-}}\left(
-X(k_{y})\right) $ 
and $\alpha $ is obtained from the normalization of the wave functions. 
Recall that the dependence on energy $E=\left( \nu +1/2\right) \hbar \omega
_{B}$ manifests itself via the functions $\mu _{\pm }\left( \nu ,\gamma
\right) $ and $c_{\pm }\left( \nu ,\gamma \right) $, given by Eqs.$~\left( %
\ref{eqn15}\right) $ and $\left( \ref{eqn16}\right) $. The dispersion
relation $(\ref{eqn23})$ is quadratic with respect to the parabolic
cylindric functions, therefore for a given band index, $n$, it has two
solutions, $E_{sn}\left( k_{y}\right) $, where $s=\uparrow $ and $\downarrow 
$ corresponds to the spin edge states with the two spinor wave functions 
\begin{equation}
\psi _{k_{y}}^{\uparrow ,\downarrow }(\xi )=\left. \psi _{k_{y}}(\xi
)\right\vert _{E=E_{\uparrow ,\downarrow n}\left( k_{y}\right) }~.
\label{eqn}
\end{equation}

\section{Energy spectrum of spin edge states and spin current}

Further we carry out the actual calculations of the spectrum of spin edge
states, the average components of spins, and the spin current components,
carried by the skipping orbits along the edges of 2DES. 
In the presence of a perpendicular magnetic field, the efficiency of SOI is
determined by the dimensionless coupling constant $\gamma $, which is
inversely proportional to $\sqrt{B_{0}}$. Therefore, the SOI effect is
significant in weak magnetic fields unless temperature fluctuations smear
magnetic quantization effects. In such fields the Zeeman effect is small~%
\cite{kramer} and we do not consider it here. 
We carry out the actual calculations for $B_0$ corresponding to the
cyclotron splitting about $5$~K. In GaAs with the electron effective mass $%
m^{\ast }=0.067m_{0}$ such a cyclotron splitting is achieved for $B_0\approx
0.25$ T and taking the Rashba coupling $\alpha _{R}\approx 4.72$~meV~\AA ,
we have $\gamma =0.03$. For such a coupling the spin-orbit effects in the
edge state spectrum should be hardly visible. The situation is favorable in
InAs where the Rashba coupling is larger, $\alpha _{R}\approx 112.49$~meV~%
\AA\ \cite{fmesz}. Despite the smaller effective mass, $m^{\ast }=0.026m_{0}$%
, in weak fields about $B_{0}=0.1$ T, we have $\gamma =0.45$. As we see
below such a coupling results in essential modifications in the spectrum and
transport of spin edge state, measurable in experiment.

\begin{figure*}[t]
\centering
\includegraphics[width=.32\linewidth]{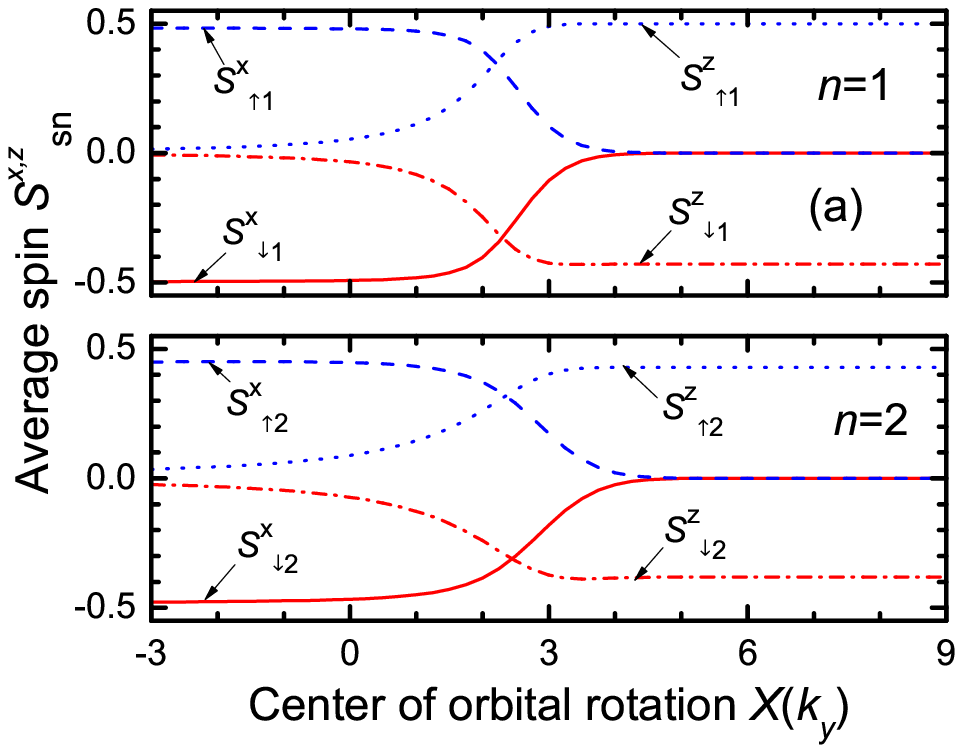} \hfill %
\includegraphics[width=.32\linewidth]{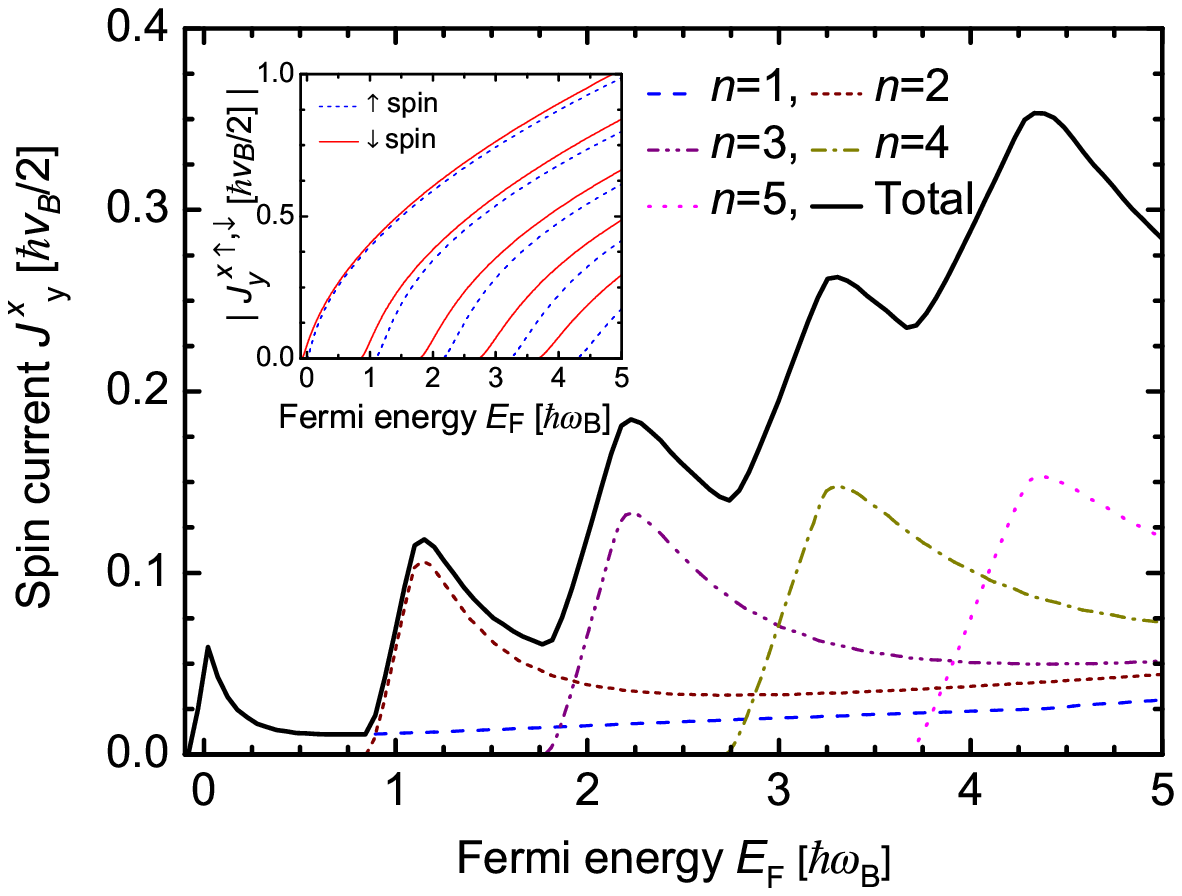} \hfill %
\includegraphics[width=.32\linewidth]{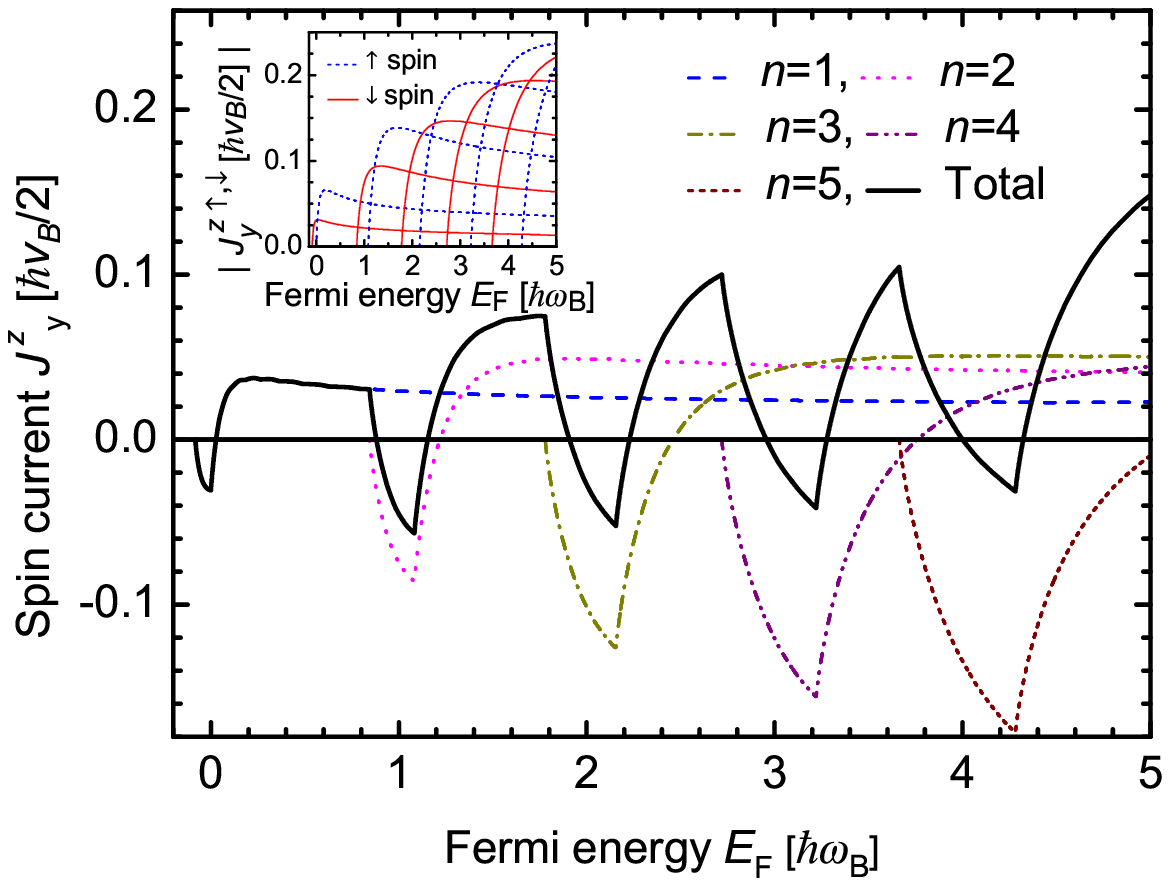}
\caption{(Color online) (a) The $x$ and $z-$components of average spins\ in
units of $\hbar $ as a function of $X(k_{y})$ for the first two bands $n=1,2$
and for $\protect\gamma =0.3$. (b) The $x-$component and (c) the $z-$%
component of the net spin current (solid curve) versus the Fermi energy. The
dashed and dotted curves plot the separate contributions to the spin
current, made by the spin up and down states together for each band $n$.
Insets show the absolute value of the separate up and down spin current
contributions.}
\label{fig2}
\end{figure*}

In Fig.~1a we plot the energy spectrum of spin edge states, $E_{sn}\left(
k_{y}\right) $, as a function of momentum $k_{y}$, which we obtain by
solving the dispersion equations ($\ref{eqn23})$. 
It is seen that for a given quantum number $n$ there are two spin resolved
magnetic edge states, $E_{\downarrow n}\left( k_{y}\right) $ and $%
E_{\uparrow n}\left( k_{y}\right) $. Both branches show monotonic behavior
in the whole range of $k_{y}$ variation. For large positive values of $k_{y}$%
, the energy of spin edge states is given by the spin-split quasibulk Landau
levels, renormalized by SOI, while at negative values of $k_{y}$ the
spectrum describes the current carrying edge channels. The spin splitting of
edge states increases with the main quantum number $n$. At the stronger
effective SOI coupling the spectrum shows well pronounced anticrossings. The
development of the anticrossings can be traced clearly in Fig.~1b where we
calculate the energy of spin edge states versus $\gamma $ or, what is the
same, versus $1/\sqrt{B_{0}}$. These anticrossings in the energy spectrum
result in additional structures of the spin current versus the Fermi energy.
In Fig.~1c we plot the average transverse position of the spin edge states
from the boundary of 2DES as a function of their center of orbital motion,
defined as%
\begin{equation}
\Delta x_{sn}\left( k_{y}\right) =\int_{0}^{\infty }dxx\left\vert \psi
_{k_{y}}(x-X(k_{y}))\right\vert _{E=E_{sn}\left( k_{y}\right) }^{2}~.
\label{eqn26}
\end{equation}%
It is seen that except for large positive values of $k_{y}$, the position of
skipping orbits takes \textit{spin resolved} values so that the up and down
spin edge states are separated in space. Notice this effect of the SOI
induced \textit{spatial separation} is missed in the approximation, adopted
in Ref.~\onlinecite{pala}. The differences of the probability density for
different spins and wave vectors of the first two bands are clearly shown in
the insets. As seen from the lower inset, the probability density for
different bands and spins differs even at large positive values of $k_{y}$.
In this limit, however, irrespective the quantum number $n$ and the spin
orientation, the particle average thickness is the same and varies linearly
with its center of orbital motion. This is because in the quasibulk Landau
states far from the interface, electrons oscillate symmetrically with
respect to the guiding center, $X(k_y)$, independent of the spin and band
index.

From the obtained spectrum we calculate the corresponding group velocities
along $y-$direction, $v_{sn}(k_{y})=\partial E_{sn}(k_{y})/\partial \hbar
k_{y}$ ($v_{x}\equiv 0$ along $x-$direction) as well as the average spin
components along $x,z-$directions 
\begin{equation}
S_{sn}^{x,z}\left( k_{y}\right) =\frac{\hbar }{2}\left. \int_{0}^{\infty
}dx\psi _{k_{y}}^{\dag }(x)\hat{\sigma}_{x,z}\psi _{k_{y}}(x)\right\vert
_{E=E_{sn}\left( k_{y}\right) }~.
\end{equation}%
Because the transverse wave functions are real, the $y-$component of the
spin vanishes identically, $S_{sn}^{y}\left( k_{y}\right) \equiv 0$. In
Fig.~2a we plot $S_{sn}^{x,z}\left( k_{y}\right) $ as a function of $%
X(k_{y}) $ for the first two bands. At large positive values of $k_{y}$ when
electrons are far from the hard-wall, the spins are mainly aligned along $z-$%
axes. This is because in the quasibulk Landau states electrons have no
preferential direction in the $(x,y)-$plane of their cyclotron rotation. In
the opposite limit of negative $k_{y}$, the edge channels are formed and the
spins are mainly aligned in $x-$direction, perpendicular to the $y-$%
direction of electron propagation. Notice that due to the spin splitting the
absolute values of the average spin components do not equal in the up and
down states and this asymmetry becomes stronger with the band index $n$.

In Fig~$2$b and $2$c we calculate the $x$ and $z-$components of the net spin
current along $y-$direction, defined as 
\begin{equation}
J_{y}^{x,z}(E_{F})=\sum_{s,n}\left. S_{sn}^{x,z}\left( k_{y}\right)
v_{sn}\left( k_{y}\right) \right\vert _{E_{sn}\left( k_{y}\right) =E_{F}}~.
\label{eqn25}
\end{equation}%
We use this definition of the spin current, which is widely accepted and
intuitive from the physical point of view. Notice that according to some
recent suggestions$~$\cite{tokalty,sun}, the standard definition of spin
current (which is the one we use here) is a proper definition and there is
no need for other definitions$~$\cite{vernes,xing,niu}.\textbf{\ }It is seen
that $J_{y}^{x}$ exhibits regular peaks as a function of the Fermi energy.
Such an oscillatory behavior is due to the SOI induced splitting in the edge
state spectrum. For a given quantum number $n$ the splitting always creates
a narrow energy region near the peak position (see the inset in Fig~$1$b)
where only the $\downarrow $ spin states contribute positively to the spin
current. With an increase of the Fermi energy the $\uparrow $ spin edge
state starts to contribute negatively at the position of $\uparrow $ spin
Landau level. In this region the exponential increase of the velocity~\cite%
{smb5} of the $\uparrow $ spin edge state with the energy results in a sharp
peak of the spin current. With a further increase of the Fermi energy the $%
\downarrow $ and $\uparrow $ spin edge states of the next $n+1$ band start
to contribute similarly but with stronger amplitudes because of the
spin-splitting enhancement with $n$. Thus the peaks of the spin current are
imposed against the monotonic background and have a period, determined by
the cyclotron energy. The latter is renormalized in the presence of SOI, as
seen in Fig.~1a in the limit of $k_{y}\rightarrow \infty $.

As seen in Fig.~2c the $z-$spin current $J_{y}^{z}$ changes its sign, in
addition to its peaked behavior: due to an interplay between the average
spin $S_{sn}^{z}\left( k_{y}\right) $ and the velocity $v_{sn}\left(
k_{y}\right) $, the spin current is negative near the Landau levels and
positive between them. In this case, at high energies corresponding to large
negative values of $k_{y}$ in each band $n$ (cf. Fig.~1a), the large values
of $v_{sn}\left( k_{y}\right) $ are compensated by the small values of $%
S_{sn}^{z}\left( k_{y}\right) $ (cf. Fig.~2a). Therefore the overall
monotonic increase of $J_{y}^{z}$ with $E_{F}$ becomes less pronounced.


\section{Summary}

In conclusion, we present an exact analytical solution to the spin edge
states, induced by the combined effect of the Rashba SOI and of the
hard-wall confining potential. The exact solution of the problem allows its
deeper intuitive understanding and can be a strong input in studying the
spin transport through edge channels. We find that due to SOI the spin edge
states are resolved not only in the energy but are also separated in space.
In the bulk of sample the electron spin is mainly aligned along the magnetic
field while near the hard-wall the spin becomes mainly perpendicular to the
edge state propagation direction. The magnitude of spin components is
asymmetric in the up and down states. We show that the spin current
components exhibit oscillations versus the Fermi energy. The predicted
oscillations, with a period determined by the renormalized cyclotron energy,
can serve as an effective tool to control the spin motion in spintronic
devices.

\section{Acknowledgements}

We thank J. Fabian and G. Vignale for useful discussions and acknowledge
support from the Volkswagen Foundation, EU under Grant PIIF-GA-2009-235394,
SFB under Grant 689, and ANSEF under Grant PS-1576.

\end{document}